# Scientific Research as a Weapon in Russia's Hybrid War in Europe: an Example of the Joint Institute for Nuclear Research in Dubna, Russia

**Dr. Tetiana Hryn'ova (CNRS, France)**

**Senior Researcher at CNRS, France**

**Abstract.** This paper examines how the Joint Institute for Nuclear Research (JINR), an international organization formally committed to peaceful science, is deeply embedded in an ecosystem of military-industrial enterprises in the city of Dubna in Russia, contributing to training specialists and developing technologies used in Russia's military operations, including attacks on civilian facilities in Ukraine. It also shows how JINR collaborates with scientific institutions on the Ukrainian territories occupied by Russia, legitimizing the occupation and exposing international partners to legal and ethical risks. Despite these ties, JINR maintains broad international collaborations, allowing its scientists and engineers to access advanced technologies and indirectly support Russia's military capabilities, highlighting the need for greater awareness in the global scientific community and coordinated sanctions enforcement.

**Résumé.** Cet article examine comment l'Institut unifié de recherche nucléaire (Joint Institute for Nuclear Research – JINR), une organisation internationale officiellement consacrée à la recherche scientifique pacifique, est profondément intégré dans un écosystème d'entreprises du complexe militaro-industriel dans la ville de Doubna en Russie. Il montre que cet institut contribue à la formation de spécialistes et au développement de technologies utilisées dans les opérations militaires russes, y compris lors d'attaques contre des infrastructures civiles en Ukraine. L'étude met également en évidence la collaboration du JINR avec des institutions scientifiques situées dans les territoires ukrainiens occupés par la Russie, ce qui contribue à légitimer l'occupation et expose les partenaires internationaux à des risques juridiques et éthiques. Malgré ces liens, le JINR continue d'entretenir de vastes collaborations scientifiques internationales, permettant à ses chercheurs et ingénieurs d'accéder à des technologies avancées et de soutenir indirectement les capacités militaires russes. Ces dynamiques soulignent la nécessité d'une plus grande vigilance au sein de la communauté scientifique internationale ainsi que d'une coordination accrue dans l'application des régimes de sanctions.

**Mots clés:** Complexe militaro-industriel (CMI), Recherche à double usage, Science et sécurité, Coopération scientifique internationale, Contournement des sanctions, Institut unifié de recherche nucléaire (JINR), Guerre Russie–Ukraine

**Acknowledgments**
Previous versions of this paper were presented at a seminar at the University of Lausanne on October 9, 2025, and during an international conference "War at Europe's Doorstep: Geopolitical, Economic, and Social Implications of the Russian-Ukrainian War" on November 6-7, 2025, also at the University of Lausanne. I am grateful to the participants of these events for their stimulating feedback on my work.

# Table des Matières/Contents

# Introduction

Russia's war on Ukraine prompted multiple sanctions against research institutes and universities in Russia [Kozmenko, 2025]. While many of these restrictions build upon earlier measures imposed by the United States in 2014 [Milov, 2025], European countries have taken the lead in expanding sanctions since 2022. By the end of 2025, the European Union had approved nineteen packages of sanctions across multiple sectors, including restrictions affecting scientific cooperation and research partnerships [EU, 2025]. These measures reflect a growing recognition that scientific institutions in Russia are embedded within a broader technological and industrial infrastructure inherited from the Soviet period. Unlike many Western academic institutions, Soviet and post-Soviet research institutes often retained substantial manufacturing capacities, enabling them not only to conduct experimental research but also to produce specialized equipment, including technologies relevant to the Russian military-industrial complex (MIC).

Technological innovation within Soviet research institutes and MIC enterprises was frequently driven by intelligence-gathered information and state-directed technological development strategies [Zhuk, 2024, p. 194]. After the dissolution of the Soviet Union in 1991, institutional boundaries between the military-industrial sector, security agencies, and academic institutions became increasingly porous. Personnel mobility among these sectors, combined with limited resources and the need to sustain research capabilities, fostered extensive cooperation through shared training programs, joint ventures, and technological partnerships [Hosaka, 2023]. While recent scholarship has examined the militarization of Russian academia [Alava, 2025] and its links with intelligence agencies [Hosaka, 2025], comparatively little attention has been devoted to the connections between academic research institutions and the broader military-industrial ecosystem. Addressing this gap is the central aim of the present study.

This paper focuses on Dubna, a city located approximately 110 km north of Moscow and home to the Joint Institute for Nuclear Research (JINR), an international organization established in 1956 to conduct fundamental research in nuclear science. JINR originated from one of the Soviet Union's classified nuclear laboratories and was conceived as what contemporaries described as "the socialist analog of the European Organization for Nuclear Research (CERN)," founded two years earlier near Geneva on the Franco-Swiss border [IUPAP, 2024, p. 183]. The establishment of both the city of Dubna and JINR itself was closely connected to Soviet state security institutions, with the NKVD (People's Commissariat for Internal Affairs) playing a significant role in their creation [Prokh, 2022, pp. 279–281]. This historical background illustrates the longstanding entanglement between scientific infrastructure and state security structures within the Soviet research system.

The sections that follow examine Dubna's industrial enterprises, JINR's collaborations, and the institute's research and educational activities to explain why, despite the JINR Charter stipulating that "the research results obtained at the Institute can be used only for peaceful purposes for the benefit of all humankind" [JINR, 1999], the institute was placed under Ukrainian government sanctions on 8 August 2025 [UA JINR; OS JINR]. More broadly, the article argues that JINR functions as a key interface between international scientific collaboration and the Russian military-industrial ecosystem in Dubna.

# 1. Literature Review and Methodological Approach

Since the foundation of JINR, Dubna has become widely known as the "open" [Sissakian, 1999], "international" [Starchenko, 2006], and "science city" [Britannica].

It had a population of 74032 people in 2024 [Rosstat, 2024]. JINR is located in Dubna on the right bank of the Volga River, while the left bank hosts "the second most classified plant within the Russian military industry" [Rhodus, p. 51] - *Raduga Missile Design Bureau* (Raduga)*, its production facility, the Dubna Machine-Building Plant* (DMZ), and other enterprises connected with Russia's MIC.

Although information on the activities of MIC enterprises is typically classified, Russia's invasion of Ukraine has heightened interest in the organization of their operations and extensive research on the subject. Recent studies [Rhodus; KSE] indicate that the Russian MIC is dominated by a few major corporate groups, including Rostec, Tactical Missile Corporation (TMC), Roscosmos, and Rosatom. Rostec, a conglomerate comprising over 800 companies, is responsible for the production of most of Russia's land-based, aerial, and naval weapons. TMC manufactures the majority of Russian cruise and anti-ship missiles, as well as guided bombs. Roscosmos, Russia's state corporation for space activities, oversees a vast network of enterprises encompassing the country's civilian and military space activities. Rosatom, Russia's State Atomic Energy Corporation, manages a broad network of subsidiaries involved in civilian nuclear programs domestically and abroad, as well as the management of Russia's military nuclear capabilities. The ownership structures and interrelations among many of these MIC groups remain opaque [Rhodus; KSE]. For example, in 2022, the Public Joint-Stock Financial Corporation AFK Sistema publicly divested from its MIC subsidiaries, including the Kronshtadt Group, which produces drones and other aerospace equipment [Sistema, 2022]. Nevertheless, AFK Sistema Group continues to pursue projects on drone engine development in collaboration with JINR [JINR, 2023, p. 4].

Since 1991, Russian MIC enterprises have expanded their public communication efforts to increase their visibility, improve their public image, and attract workers in an increasingly competitive labor market. These developments are clearly observable in Dubna. During the Soviet period, the DMZ concealed its involvement in cruise-missile production by being officially presented as a leading manufacturer of baby strollers, producing 4.5 million units between 1953 and 1990 [Prokh, 2022, p. 116]. In 2016, a museum exhibition on the history of cruise-missile development at Raduga and DMZ was created at the Dubna School No. 10, openly displaying the majority of missiles produced in the city [Dubna, 2022]. In 2022, V. Prokh, the first mayor of Dubna (1991–2014), together with journalist D. Sokolov, published a book "*Science-City of Dubna: History of an Eternal City*" [Prokh, 2022], which offers the first publicly available account of the activities of MIC enterprises in the Dubna area.

In the early 1990s, the Dubna city administration, the management of JINR, Raduga, DMZ, and the other MIC enterprises in the Dubna area formed the Board of Directors of city-forming enterprises of Dubna [Prokh, p. 62]. This board reinforced links between these institutions through coordinated investments in the development of the city infrastructure and public activities in the Dubna area. In the past 35 years, these links were strengthened through JINR educational and knowledge transfer programs, described by B. Starchenko, then Scientific Secretary of JINR: "In the late 1990s, the concept of JINR as a large multidisciplinary international centre for fundamental research in nuclear physics and related fields of science and technology was adopted. The aim is to transfer the results of highly technological research at JINR to applications in industrial, medical, and other technical areas, to provide additional sources of financing for fundamental research and the organization of new working places for specialists who are involved with these broader topics at the institute." [Starchenko, 2006]. The collaboration between enterprises in Dubna and JINR, as well as JINR's broader research activities, is documented in the annual public *Topical Plans for JINR Research and International Cooperation* [JINR TP], called *research plans* thereafter. These research plans are produced at the end of each calendar year and are approved at the last JINR Scientific Council of the year. They have been available online in both English and Russian since 2007, detailing JINR projects for each calendar year and listing the institutions involved. The Russian-language version additionally provides contact persons for each participating institution. In addition to information obtained from these research plans, the material presented in this paper draws on news items published on the websites of JINR, the Dubna city administration, and other Dubna-based enterprises, as well as JINR annual reports and knowledge-transfer brochures. The OpenSanctions database [OS] is used to provide information and references regarding the sanctions status of the institutions and enterprises discussed. Methodologically, the study relies on qualitative analysis of institutional documents, publicly available research plans, open-source intelligence materials, and sanctions databases. These sources are triangulated to reconstruct the network of institutional and technological connections linking JINR with enterprises of the Russian military-industrial complex.

The remainder of the paper is structured as follows. The next two sections provide a concise history of Raduga and its production facility in Dubna and an overview of the major enterprises of Dubna, primarily based on [Prokh, 2022]. Section 5 outlines the educational programs in Dubna. Sections 6-8 examine JINR's knowledge-transfer initiatives and its collaborations with sanctioned enterprises and institutions in Dubna and outside it. Section 9 reviews JINR activities on the Ukrainian territories currently occupied by Russia. Section 10 describes JINR's ongoing international collaborations. Finally, Section 11 summarizes how JINR links the international scientific community to the Russian MIC enterprises, allowing them to circumvent the ongoing sanctions. The case of JINR can also be interpreted through the analytical lens of dual-use research and the growing literature on the geopolitics of science, which highlights how scientific infrastructures and international collaborations can become embedded in geopolitical competition and security strategies.

## 2. Dubna as a Soviet and Post-Soviet Military–Industrial Cluster

The origins of Dubna's military–industrial complex and the establishment of the DMZ can be traced to the construction of the Moscow–Volga Canal in the late 1930s and the construction on its banks of a plant for the serial production of the MTB-2 heavy seaplane bomber, developed by the A.N. Tupolev Design Bureau [Prokh, 2022, p. 104]. A major boost for the development of the left-bank industrial site in Dubna came from German reparations following World War II: in 1946, 500 specialists from design bureaus in Dessau and Halle arrived with their families, along with equipment and machinery from the Arado aircraft factory in Warnemünde [Prokh, 2022, p. 108]. These reparations furnished machine shops, an aerodynamic laboratory, an engine laboratory, a hydro-laboratory, a measurement laboratory, and other facilities in Dubna. Two design bureaus were subsequently established: OKB-1, focused on the design and development of heavy turbojet bombers and headed by Brunolf Baade, with the Soviet engineer P. N. Obrubov as his deputy; and OKB-2, devoted to the design and development of experimental aircraft with liquid-propellant rocket engines, headed by Hans Rössing, with A. Y. Bereznyak as his deputy. The German specialists departed in 1951–52, and the DMZ was assigned a new mission: mastering the serial production of cruise missiles [Prokh 2022, p. 111, 135]. In 1951, a specialized design department was established at the plant as a branch of A.I. Mikoyan's OKB-155. This branch became an independent machine-building design bureau Raduga in 1957. A. Y. Bereznyak initially headed the branch and subsequently led the design bureau, which today bears his name.

Over the years, specialists at the DMZ, in collaboration with designers from Raduga, organized the production of cruise missiles, including the KS, P-7, K-10, Kh-20M, P-15, KSR-2, KSR-11, Kh-28 series, the Kh-22 family, Kh-58, Kh-59M, Kh-55, Moskit, Kh-15, Kh-101, and others. The plant also developed expertise in aviation technology, producing fuselages, wings, vertical stabilizers, pylons, and nose cones for the MiG-25 supersonic fighter jet [Prokh, 2022, p. 111; NTI 2002].

In 2004, Raduga became part of the TMC [Prokh, 2022, p. 164] and the DMZ became part of the AFK Sistema Group [Prokh, 2002, p. 118]. At this point, the DMZ received a major government contract to produce, repair, and modernize tactical missile weapons and equipment for military aviation, which accounted for up to 75% of its total production volume. Simultaneously, serial production of a wide range of satellite communications antennas was established. At the request of JSC RAC MiG, the plant's specialists mastered the production of add-on fuel tanks and external refueling units for various MiG-29 aircraft modifications. The plant produces beam support pylons, cockpit components, and assemblies for the Su-25 for Sukhoi LLC. A contract for the production of suspended electronic warfare (EW) pods for various Su-25 aircraft modifications was awarded in 2011." [Prokh, 2022, p. 121]. In 2020, the DMZ plant became part of the Kronstadt Group, a subsidiary of AFK Sistema. Kronstadt also established a separate facility, the Kronstadt drone plant in Dubna, adjacent to the DMZ site, for the serial production of large unmanned aerial systems, including the *Orion* and *Inokhodets* models. DMZ's role in these drone-production activities includes manufacturing ground control stations, transport containers, tooling, components, and various parts [Prokh, 2022, p. 122]. Over time, Raduga and DMZ became central actors in the Soviet and later Russian cruise missile production system, contributing to multiple generations of missile platforms used by the Russian armed forces.

Russia's full-scale invasion of Ukraine in February 2022 led to the extensive use of weapons produced by DMZ, Raduga, and the Kronstadt drone plant in Dubna against civilian targets in Ukraine. Notable examples include:

-   A Kh-22 missile strike on the Amstor shopping center in Kremenchuk on 27 June 2022, which killed at least 20 people and injured about 60 others [ECHO, 2024];

-   A Kh-22 missile strike on a residential high-rise in Dnipro on 14 January 2023, resulting in at least 46 fatalities [O'Grady, 2023];

-   A Kh-101 missile strike on the Okhmatdyt Children's Hospital in Kyiv on 8 July 2024, when 627 young patients were inside the facility; two medical workers were killed, and at least eight children were injured [MFA, 2025].

Although these three Dubna enterprises are currently under sanctions [OS DMZ; OS Raduga; OS Kronstadt], they continue to expand their production capacities [ECHO, 2024]. For example, Kh-101 missile production has increased eight- to tenfold since 2022, reaching an estimated 420–600 units per year and incorporating significant technical upgrades [FT, 2024; 24TV]. These missiles rely on the availability of Western components, machine tools, and qualified personnel [Rhodus]. The expertise of Raduga's personnel has likewise facilitated the rapid development of Russia's guided bomb program, which has had severe consequences for Ukrainian defence and civilian populations [Bolgarin, 2024]. These fast-track improvements in missile and bomb design, as well as their production turnover, were made possible by the JINR's long-term cooperation with these military enterprises on personnel training as well as research and infrastructure projects in Dubna, which are discussed in Sections 5 and 7, respectively. It should be noted that sanctions are critical to slow the pace of these developments and to undermine their long-term sustainability [Litranovych, 2025].

# 3. Other Enterprises in Dubna

Raduga, DMZ, and the Kronstadt drone plant are not the only MIC enterprises in Dubna. The city is also home to a range of other industrial facilities with a long history of cooperation with the Russian Army, Navy, the Ministry of Defense and the Federal Security Service of the Russian Federation (FSB). A selection of these facilities is listed below in the order of their establishment.

The *Tensor* Instrument Plant, one of Dubna's key city-forming enterprises, was established in 1973 to produce equipment for nuclear power plants, including in-reactor monitoring systems. Since 1991, Tensor has supplied the Russian Ministry of Defense with integrated physical and fire protection systems. This cooperation began as part of the Nunn-Lugar Cooperative Threat Reduction program, aimed at securing major chemical and nuclear weapon facilities. As presented by the director of Tensor in 2022, "Future projects involve modernizing physical security systems at particularly hazardous oil and gas facilities, including refineries and fields." [Prokh, 2022, p. 212].

The *Atoll* Research Institute was established in 1976 to provide the equipment for monitoring underwater and surface conditions and is currently working with the Russian Navy to secure the borders of zones of economic and strategic interest [Prokh, 2022, pp. 223-235].

The *Space Communication Center* (SCC), another of Dubna's key city-forming enterprises, was established in 1977 and is part of the Russian Satellite Communications Company (RSCC). It remains the largest facility of its kind in Russia and one of the largest in Europe. In 1999, during the Second Russian–Chechen War, Dubna SCC provided communications support for Russian military units [Prokh, 2022, p. 268]. During the Russo-Georgian War in August 2008, it blocked Georgian television channels in the Georgian region of South Ossetia to ensure fully pro-Russian coverage of the conflict [Prokh, 2022, p. 269]. Likewise, in March 2014, during the Russian invasion of Ukraine, the Dubna SCC blocked all Ukrainian channels in occupied Crimea to guarantee that television broadcasting was fully controlled by the Russian government before Russia's illegal referendum there [Prokh, 2022, p. 270]. Despite its central role in Russia's military and information operations, Dubna SCC continued its technical cooperation with Eutelsat [Prokh, 2022, p. 273], one of Europe's major satellite operators. The acquired experience "has been successfully applied to similar projects for control and monitoring of other satellites, including RSCC's own fleet." [SpaceNews, 2011]. In the early 2000s, Dubna SCC's fiber-optic channels were also used to support data transfer between JINR and leading scientific centers worldwide, including CERN on the Franco-Swiss border, FNAL and BNL in the USA, and DESY in Germany [Budagov, 2007; Prokh, 2022, p. 272]. The Dubna SCC remains involved in JINR's collaborative projects with CERN [JINR TP, 2026, pp. 27–33].

V. Prokh writes [Prokh, 2022, p. 297-298] that in the late 1980s, "Almost the entire working population of the city worked at the five city-forming enterprises or in the organizations that serviced them. … But the events of the late 1980s and early 1990s left the city-forming enterprises and scientific institutes virtually completely without orders. While the management of these organizations sought ways and means to save them, many scientists and engineers decided to try their hand at small business, and, it must be said, with considerable success. While the country was transitioning from a planned to a market economy, by 1992, two thousand cooperatives and small businesses were already operating in Dubna. Moreover, a large number of them specialized in high-tech products."

This period was characterized by privatization, reorganization, and the establishment of new enterprises, often building upon preexisting facilities. For instance, the Research Institute of Applied Acoustics (NIIPA) in Dubna, officially established in 1994, celebrated its 75th anniversary in 2025 [Sadkevitch, 2025], reflecting its historical roots in Soviet-era acoustic research in Dubna. NIIPA conducts scientific investigations in acoustics and hydroacoustics, including the development of acoustic test stands to assess the effects of acoustic vibrations on rocket equipment. In 2005, the institute was placed under the authority of the Federal Service for Technical and Export Control of Russia, broadening its mandate to encompass research supporting federal technical and export control functions. Currently, NIIPA is subject to sanctions imposed by multiple countries [OS NIIPA].

Many new enterprises were formed by JINR scientists and were initially located on the JINR territory [Prokh, 2022, p. 302], e.g., the *ASPECT Research and Production Center, Dedal Research and Production Complex, DVIN LLC*, to name a few. ASPECT, founded by JINR scientists and Tensor engineers in 1991, specializes in radiation monitoring equipment of various complexity, "from multilayer printed circuit boards to materials for scintillation gamma and neutron detectors" [ASPECT]. DVIN, founded jointly by JINR and the FSB in 2008, develops explosive and narcotics detectors [DVIN]. *Dedal,* currently part of the Rosatom State Corporation and under sanctions by Ukraine [OS Dedal], develops and manufactures devices for physical protection systems of critical and high-risk civilian and military facilities.

These entrepreneurial activities naturally led to the idea of creating a technology and innovation cluster around the JINR, with the intent to leverage its scientific expertise and infrastructure to support industrial and technological growth in the region [JINR INNOVATIONS, Starchenko, 2006]. The Dubna *Special Economic Zone* (SEZ) was created in 2005 [SEZ], focusing on high technology and nuclear/information sectors. Many of the over 170 SEZ enterprises have been under sanctions since 2022 [OS Dubna, SVIT, 2024]. In 2025, the 19th package of EU sanctions specifically targeted Russia's SEZs, including the Dubna SEZ, due to their "critical role in driving economic growth and infrastructure development" [EU, 2025]. But the scope of these sanctions is quite limited: it bans the EU businesses from entering into *new* contracts with any entity established within the SEZ. It does not prevent the SEZ activities of JINR scientists working on projects within the EU, for example.

# 4. Educational Programs and Workforce Reproduction

In addition to the institutes and enterprises discussed in previous sections, in 1973-1993, Dubna city hosted the Volzhsky Higher Military Construction Command School, also known as "Military School of the USSR Atomic Project" [JINR Mag, 2023], which was moved there from Novosibirsk. As Prokh writes, "The new location of the educational institution was successful, since the school's teaching staff was replenished with highly qualified personnel from scientific institutions in the city. " [Prokh, 2022, p. 290]. The cadets took part in the construction of facilities for "the Joint Institute for Nuclear Research, the Tensor plant, the Atoll Research Institute, the Kurchatov Institute in Moscow, Protvino, Akademgorodok, and enterprises in Novosibirsk, among others" [Prokh, 2022, p. 293].

After the Volzhsky Higher Military Construction Command School was closed, its facilities were transferred to the city and repurposed to house Dubna State University (DSU). "In 1994, on the initiative of the JINR directorate, and with the active support of the Russian Academy of Natural Sciences, the town of Dubna and the Moscow region administrations established the Dubna International University of Nature, Society and Man. There are dozens of JINR staff members, all renowned scientists, among the university staff. The university educational base is actively developed on the territory of JINR, so that Dubna has become a town of students as well as physicists," wrote the JINR Scientific Secretary, Boris Starchenko, in an article dedicated to the JINR 50th anniversary [Starchenko, 2006]. DSU educational programs are specifically targeted to train personnel for the enterprises in the Dubna area [Prokh, 2022, p. 369] and are taught by JINR scientists [Fursaev, 2009, p. 30]. JINR scientist and then- DSU-rector D. Fursaev, in his interview with RIA News in 2021 said that "Such educational projects are joint investments between the university and JINR" and that "the institute is assisting in the development of relevant university infrastructure and funding several other expenses." [RIA, 2021]. In 2009, 78 DSU graduates were employed in the SEZ at 14 resident enterprises [Fursaev, 2009, p. 26]. Table 1 shows that over 30% of DSU graduates joined MIC enterprises in the Dubna area. This conservative number from 2009 does not take into account more recent dedicated DSU programs for the Dubna SCC, which has trained 10-12 engineers per year since 2017 [Prokh, 2022, p. 272], and a new 2-year program launched in 2023 to train 30 engineers per year for the Kronstadt drone production plant [KR, 2023].

**Table 1.** Distribution of DSU graduates in Dubna-area enterprises in 2009. From [Fursaev, 2009, p. 30].

| Enterprises | Hired of DSU graduates per year |
|---|---|
| **JINR** | 30% |
| **Raduga** | 10% |
| **Tensor** | 10% |
| **Atoll** | 6% |
| **DMZ, ASPECT, NIIPA** | 13% |

In addition to the DSU programs above, in 1999-2015, JINR scientists taught a course on "Electronics of physics facilities" [JINR TP, 2015, pp. 176-178] on the Dubna campus of the Moscow State Institute of Radio Engineering, Electronics and Automation (MIREA) to train specialists for the JINR and Raduga [JINR, 2009]. Since 1991, JINR has also run dedicated education programs in the JINR-based departments of Russian universities [JINR EDU]. Over 40 JINR scientists teach in these programs [Fursaev 2009, p. 30]. All four universities with JINR-based departments are currently under sanctions, as described below. Moscow Institute of Physics and Technology (MIPT) is under sanctions by the EU, UK,  USA, Canada, Switzerland, Japan, Ukraine, and New Zealand [OS MIPT] because of its development of drones and fighter aircraft equipment for Russia's military. MIPT is part of a consortium of Russian institutions involved in training specialists for Russia's MIC. National Research Nuclear University "MEPhi" [OS MEPhi] and St. Petersburg State University [OS StPbSU] are under sanctions by Ukraine. While M.V. Lomonosov Moscow State University is under sanctions by the USA and Ukraine [OS MSU]. Its Dubna campus was inaugurated in 2022. In addition to the university-level programs outlined above, JINR also runs educational programs for school teachers in Dubna and, between 2009 and 2019, at CERN [JINR TP; CERN RTP].

As illustrated above, over the past 30 years, the JINR staff successfully trained personnel for various MIC enterprises described in Sections 3 and 4 through educational programs at DSU and other universities in Russia. This well-trained personnel is critical for Russia's current military effort in Ukraine as well as its hybrid war activities in Europe, because modern warfare depends on highly skilled specialists who can design, operate, maintain, and modernize advanced weapons systems, electronic warfare tools, communications infrastructure, aerospace technologies, and defense-sector production lines, all of which require expertise that cannot be substituted by unskilled labor.

# 5. JINR Knowledge Transfer Programs

In addition to the educational programs discussed in the previous section, JINR has accumulated extensive experience in knowledge-transfer initiatives, particularly since the creation of the Special Economic Zone (SEZ) in Dubna in the mid-2000s. Within this framework, the institute increasingly positioned itself not only as a center of fundamental scientific research but also as a platform for technological commercialization and industrial collaboration. Knowledge transfer plays a central role in this strategy by enabling the technological outputs of fundamental research to be transformed into practical applications across a variety of industrial sectors. In the Dubna innovation ecosystem, this process typically occurs through collaborative research projects, joint development initiatives, and the creation of spin-off companies that translate laboratory technologies into industrial products. Several examples of such ongoing projects, including those with potential dual-use applications, are documented in the JINR *Big Science – Business* brochure [JINR, 2023].

One of the most notable projects described in this document involves collaboration with Hydrogen Technology Center LLC, a company affiliated with the AFK Sistema Group. The goal of this initiative is to develop a helicopter-type unmanned aerial vehicle (UAV) powered by hydrogen fuel cells [JINR, 2023, p. 4]. Although such technologies are often presented in the context of civilian applications—such as environmental monitoring, infrastructure inspection, or emergency services—fuel-cell UAV platforms also possess clear potential for military and security uses. Their extended endurance, reduced thermal signature, and operational flexibility make them particularly attractive for reconnaissance missions and other strategic tasks. Consequently, drone-development research of this type is of considerable interest to other Dubna enterprises, including the Kronstadt Group, which specializes in unmanned aerial systems. The significance of these technological developments was publicly emphasized during a meeting of the Council for Science and Education of the Russian Federation, chaired by President Vladimir Putin at JINR on June 13, 2024, where UAV-research projects were highlighted as an example of the successful industrial application of fundamental scientific research [Putin, 2024].

Another example of JINR's knowledge-transfer activities is its collaboration with BRS LLC, a company affiliated with the Rostec State Corporation, one of the largest conglomerates within Russia's military-industrial complex. In this project, JINR researchers contribute their expertise in detector technologies by developing a hybrid pixel detector that combines a GaAs:Cr or CdTe semiconductor sensor with a custom-designed ASIC chip [JINR, 2023, pp. 12–13]. Such detectors are widely used in high-energy physics experiments and advanced imaging systems. However, their technological characteristics, particularly high radiation tolerance and precision detection capabilities, also make them suitable for a range of security, defense, and aerospace applications. The participation of JINR in these projects illustrates how technologies originally developed for fundamental scientific research can be adapted for industrial and strategic uses.

JINR has also developed collaborative projects in the field of materials science and filtration technologies. One such initiative involves REATREK-Filter LLC, a company specializing in advanced filtration systems. In this project, the institute contributes track-membrane technologies originally developed for scientific research, which are subsequently used by REATREK to produce high-performance filtration products [JINR, 2023, p. 10]. These membranes enable the precise separation of particles and microorganisms at the microscopic level and have applications ranging from medical diagnostics to water purification systems. In particular, portable filtration devices produced by REATREK using JINR track membranes are marketed to military personnel and workers operating in extreme environments where reliable water purification systems are required [REATREK]. The use of these technologies illustrates how scientific innovations developed within physics research laboratories can be adapted to practical applications with direct relevance for military logistics and field operations.

Another area in which JINR knowledge-transfer activities intersect with strategic industries concerns aerospace engineering and space technologies. JINR scientists have collaborated with Technomash, a subsidiary of the Roscosmos State Corporation, on the development of a neutron-activation diagnostic method designed to detect hidden structural defects within the engines of Proton-M launch vehicles without disassembling them [AstroNews, 2017]. This technique represents a significant technological innovation in non-destructive testing, allowing engineers to identify structural weaknesses in complex aerospace components with a high degree of precision. Beyond its relevance for civilian space programs, such diagnostic technologies are also applicable to military rocket systems and other aerospace equipment, further illustrating the dual-use character of many JINR research outputs.

A distinctive feature of JINR's knowledge-transfer model is the central role played by intermediary firms that serve as institutional bridges between the institute and industrial partners. Many of these companies were either founded by JINR scientists or employ personnel trained at the institute, allowing them to maintain close scientific and technological connections with JINR laboratories. Through these intermediary structures, scientific knowledge developed within the institute can be transferred to industrial partners while simultaneously generating additional funding streams for research activities. In practice, this model facilitates the integration of fundamental research with industrial innovation processes in the Dubna region.

Taken together, these initiatives demonstrate that JINR's knowledge-transfer programs operate within a broader regional innovation ecosystem in which fundamental research, industrial development, and technological commercialization are closely intertwined. While these collaborations are often presented as examples of successful technology transfer and regional economic development, they also illustrate how scientific infrastructures can become embedded within industrial networks that include enterprises linked to the Russian military-industrial complex. The dual-use character of many of the technologies involved, ranging from UAV propulsion systems and radiation detectors to filtration materials and aerospace diagnostics, highlights the broader strategic implications of knowledge-transfer activities in contexts where civilian and military technological development are closely interconnected.

# 6. JINR Collaborations in the Russian Federation

The Board of Directors of City-Forming Enterprises of Dubna acts as the main platform for strategic cooperation between major Dubna enterprises and the local government. The initial board included leaders from JINR, Raduga, DMZ, Tensor, Dubna SCC, and representatives from the Dubna city administration. After establishing the DSU and Dubna SEZ, the board expanded to include its representatives. It functions as a space for aligning scientific, industrial, and educational efforts within the city and creating working groups, such as those focusing on housing or workforce policies. A typical program at Dubna companies was outlined by Tensor's director, [Prokh, 2022, p. 211]: "To retain staff at the plant, a program to support young specialists was adopted for 2017-2022. This included stipends for interns—senior students at Dubna University (under an agreement on targeted training for engineering students), interest-free loans for housing for employees with at least two years of experience, reimbursement for parents of half the cost of their child's kindergarten, and a childcare allowance for young mothers in the amount of one minimum wage for childcare from one and a half to three years. Today, social guarantees for the plant's employees are an extremely important part of JSC Tensor's development strategy." For example, in August 2021, JINR, Raduga, and Promtech enterprise of Dubna SEZ [OS Promtech Dubna] launched a joint initiative to construct new housing in Dubna [OEZ, 2021]. Such company programs are essential for attracting and retaining young specialists to work in Raduga [BIHUS, 2024] and other MIC enterprises in the Dubna area.

Beyond collaborations in personnel training, housing construction, and related activities, JINR reports also document scientific and infrastructural links between JINR and other enterprises in Dubna. One long-standing JINR infrastructure initiative, the *Multifunctional Information and Computing Complex Project*, active since at least 2003, aims to provide information, computing, and network support to several Dubna-based enterprises listed as project participants, including DMZ [JINR TP, 2012, p. 153], Raduga [JINR TP, 2016, p. 164], DSU, Dubna SCC, and Dubna SEZ [JINR TP, 2026, pp. 27–33]. In addition, JINR has used Raduga's facilities for the development and testing of scientific equipment [JINR TP, 2014, p. 51]; as reported, 'several versions of the developed and manufactured trigger-system modules were tested both at the SPS beams at CERN and at the test stands of I. P. Chupin at the MKB Raduga, under the supervision of a military representative [JINR, 2014].'

Cross-referencing JINR research plans [JINR TP] with the Open Sanctions database [OS] shows that JINR collaborates with numerous sanctioned Russian institutions outside Dubna. At least 80 of the 212 Russian partners are currently under sanctions (20 by the EU, 35 by the United States, 19 by Switzerland, seven by the United Kingdom, and 71 by Ukraine), including entities within Rosatom and the Kurchatov Institute [HRYNOVA 2024; JINR TP 2026; OS], both deeply embedded in Russia's MIC, particularly its nuclear programs.

# 7. JINR Activities in the Occupied Territories of Ukraine

After the Soviet Union dissolved in 1991, JINR, as an international organization, retained property in Ukraine, including ownership of the Dubna resort near Alushta in Crimea [Resort Dubna]. Following Russia's annexation of Crimea in 2014, JINR continued to hold conferences and other events there, regularly presenting the resort as being located in Russia in event websites, presentations, and proceedings [PPNL, 2018]. Notably, there are annual events for the JINR Association of Young Scientists and Specialists, with the most recent one held in June 2025 [Alushta, 2025]. The organization also co-sponsors student trips with DSU [Denikin, 2015]. Participants in these JINR events may face criminal and administrative liability under Ukrainian law.

Furthermore, following the Russian government's policy of scientific integration of the occupied territories of Ukraine, JINR developed collaborative projects with

- impostor institutes on the temporarily occupied territories of Ukraine: Donetsk Institute for Physics and Engineering named after A.A. Galkin [JINR TP, 2024, p. 218; JINR TP, 2026, p. 262], Donetsk State University [Zinicovscaia, 2024], Donetsk National Technical University [Zinicovscaia, 2024];

- stolen institutes on the temporarily occupied territories of Ukraine: Crimean Astrophysical Observatory in Nauchny [JINR TP, 2017, p. 204; JINR TP, 2018, p. 209] and the Institute of Biology of the Southern Seas in Sevastopol [JINR TP, 2015, p. 198].

The impostor institutes were created by Russian authorities appropriating the names and the material base from the legitimate Ukrainian scientific organizations, which moved elsewhere in Ukraine and are claiming continuity with them, while operating under occupation and without international legal recognition. In the case of stolen institutes, the legitimate Ukrainian scientific organizations did not have the possibility to move elsewhere in Ukraine. Often, JINR projects with these organizations involve international partners. For example, the JINR project in occupied Sebastopol received funding from the International Atomic Energy Agency (IAEA) in Vienna, Austria [RFERL] and is still ongoing [JINR TP, 2026, p. 267].

It should be noted that even during the Soviet times, JINR supported narratives and practices that helped to erase Ukrainian identity and appropriate Ukrainian scientific achievements. Until 1990, employment at JINR was one of the few avenues available to Soviet scientists to participate in international particle-physics research, particularly at CERN. At the same time, JINR scientists born in Ukraine, such as Vladimir Veksler, were frequently regarded as 'Russian' by their international colleagues [IUPAP, 2024, pp. 185, 187]. This perception was further reinforced in 1994, when the Russian Academy of Sciences established the Veksler Prize for outstanding achievements in accelerator physics. Such pro-Russian biases persist within the scientific community. Even in 2026, the History webpage of the International Union of Pure and Applied Physics (IUPAP) states that Russia joined the organization in 1957 [IUPAP History], although Russia did not exist as an independent state at that time; it was the Soviet Union that joined IUPAP in 1957, with physicists born in Ukraine, such as Vladimir Veksler and Abram Joffe (Ioffe), playing important role in it [IUPAP, 2024, pp. 175, 211].

JINR's continued activities in the temporarily occupied territories of Ukraine, its collaborations with impostor and stolen Ukrainian institutions, and its long-standing role in reinforcing Soviet and post-Soviet narratives that subsume Ukrainian scientific identity collectively contribute to the normalization and legitimization of Russia's control over Ukrainian territories within the international scientific community in general and its international partners in particular.

# 8. Current JINR International Research Collaborations

JINR currently has fifteen member states [JINR]: Armenia, Azerbaijan, Belarus, Bulgaria, Cuba, Arab Republic of Egypt, Georgia, Kazakhstan, North Korea (suspended in 2015 due to UN sanctions), Mongolia, Romania, Russia, Slovakia (suspended in 2023 due to Russia's invasion of Ukraine [JINR AR, 2023, p. 206]), Uzbekistan, and Vietnam. Of the original eleven founding states, only Bulgaria, Mongolia, and Romania remain active members of JINR. The main governing body of JINR is the Committee of the Plenipotentiaries of the Member-State Governments, consisting of one representative of each of the member states. Each of the member states is supposed to contribute to the JINR budget [Sissakian, 1999]. JINR budget in 2023 was close to 200 MUSD, with Russian contribution accounting for over 80%. It has about 5000 staff members [JINR AR, 2023, pp. 206, 208], of which 93% are Russian citizens [Open Dubna 2024]. In the 70-year history of JINR, only one non-Russian physicist has served as director: Prof. Dezső Kiss from Hungary, who held the position from 1989 to 1991 [CERN Courier, 1989]. JINR is actively expanding its international collaborations. Germany, Hungary, Italy, Serbia, and South Africa are listed as JINR Associate Member States, while Brazil, China, and India have recently acquired the status of JINR Partner Countries [JINR].

# Joint Institute for Nuclear Research

## Joint Institute for Nuclear Research

| Name: | Joint Institute for Nuclear Research | Total number of participants: | 327 |
| --- | --- | --- | --- |
| Country: | Russia | Users: | 244 |
| Town: | Dubna | External Participants: | 80 |
| Telephone: | +7 - 496 - 21 65 059 | Other Participants: | 3 |
| Fax: | +7 - 495 - 632 78 80 | | |

**Fig. 1.** From the CERN Experimental Program Greybook database. Image taken on 29 November 2025  [CERN Greybook].

Since 1991, JINR has become a member of UNESCO, the IUPAP, the European Physical Society, and other international organizations [Sissakian, 1999; Starchenko, 2006]. Despite the ongoing Russo-Ukrainian war and Russia's occupation of Crimea, JINR was granted observer status by the CERN Council in September 2014. This observer status was suspended in March 2022 following Russia's full-scale invasion of Ukraine, although JINR's cooperation agreement with CERN formally remains in force until January 2030, albeit with certain restrictions [Gianotti, 2024]. As shown in Figure 1, 244 JINR researchers and their family members currently retain on-site access to CERN experimental facilities in Switzerland and France, while an additional 80 researchers have access to CERN's computing environment.

Overall, JINR's international cooperation has expanded from 700 research centers and universities in 60 countries in 2006 [Starchenko, 2006] to 990 institutions across more than 70 countries in 2024–2025 (Fig. 2). However, this number does not reflect active bilateral collaborations but rather joint participation in broader international projects conducted at CERN or other global facilities, as discussed below. Many of the listed institutions are not aware that they are considered JINR "collaborators." A review of the "Alphabetical List of Collaborators" in the JINR research plans found 883 institutions listed for 2024 [JINR TP, 2024; Hrynova, 2024] and 898 for 2026 [JINR TP, 2026]. The actual number of active collaborators is likely lower, as some listed institutions do not cooperate with JINR or are located in countries that have terminated their JINR membership (Czech Republic, Moldova, Poland). It should be noted that, following Russia's full-scale invasion of Ukraine in 2022, JINR research plans include the following footnote for institutes from the Czech Republic, Poland, Bulgaria, Germany, Romania, and Slovakia: 'the cooperation may be limited by conditions adopted unilaterally by the State' [JINR TP, 2026, pp. 251, 253, 254, 260, 261, 268].

**Fig. 2** From the JINR website [JINR]. Image taken on Oct 19, 2024 [Hrynova, 2024]. Numbers confirmed to remain the same as of Dec 13, 2025.

JINR also participates in the Worldwide LHC Computing Grid (WLCG) project [Grid], which enables Dubna-based enterprises involved in the JINR Multifunctional Information and Computing Complex Project (see Section 7) to perform large-scale computations using distributed scientific computing infrastructure at research centers worldwide. Efforts by JINR to develop computing infrastructure in Dubna are supported by its international partners [Korenkov, 2023; JINR TP, 2026, pp. 27–33].

JINR is also a participant in the DarkSide dark matter experiment at the underground Gran Sasso National Laboratory in Italy [INFN], the RICOCHET experiment at ILL in Grenoble, France [CNRS], and the FAIR accelerator project in Darmstadt, Germany (where its participation is funded by Rosatom State Atomic Energy Corporation) [FAIR].

Although the U.S. government has imposed strict limitations on cooperation with Russian institutions since 2014 [Nelson 2024], the JINR research plans for 2026 [JINR TP, 2026, pp. 271–272] detail ongoing involvement in neutrino experiments at FNAL, RHIC experiments at BNL, and cooperation with 75 additional U.S. institutions – including Lawrence Livermore and Los Alamos National Laboratories, the country's two nuclear weapons design centers. Most of these collaborations seem to be due to JINR's participation in research projects at CERN.

It should be noted that scientists involved in the international collaborations described above also participate in JINR projects carried out with Rostec [JINR, 2023], the FSB [DVIN Pubs], and enterprises discussed in Sections 3–4 and 6–8. In addition, these JINR researchers teach at DSU [Alexandrov 2024] and collaborate with impostor and stolen institutions [JINR TP, 2025, pp. 144–153; Allakhverdyan, 2023], discussed in Section 9.

The JINR Charter [JINR, 1999] requires the affiliations of more than one-third of the members of the JINR Scientific Council to be from non-member states [JINR SC/PAC]. The evolution of affiliations of scientists from non-JINR-member states in the JINR governing bodies between 2024 and 2025 is presented in Table 2. It should be noted that, for the Scientific Council, most of the scientists listed in this table come from JINR Partner or Associate Member States: four from China, one from India, and one from Brazil. The remaining members come from a diverse set of countries, including Argentina, France, Israel, Mexico, South Korea, and the USA.

**Table 2** Affiliations of scientists from non-JINR member states participating in the JINR scientific council (SC) and Program advisory committees (PAC) as of October 23, 2025

(October 23, 2024 [Hrynova, 2024]). The value in superscript corresponds to the number of scientists who suspended their membership.

| | SC | PAC Particle Physics | PAC Nuclear Physics | PAC Condensed Matter |
|---|---|---|---|---|
| Associate Members | 3 (3) | 1 ($2^1$) | 2 (2) | 3 (3) |
| Other in Europe | 1 (2) | $1.5^1$ ($2.5^2$) | 1 (1) | 2 (2) |
| Other in Asia | 7 (5) | 3 (3) | 3 (4) | 1 (1) |
| Other in North America | 2 (2) | 2.5 (2.5) | 0 (0) | 0 (0) |
| Other in South America | 2 (2) | 1 (1) | 1 (1) | 0 (0) |
| CERN | 0 | $2^2$ ($2^2$) | 0 (0) | 0 (0) |

The participation of international scientists in JINR's supervisory bodies upholds its international status and reputation, even though its policies are substantively dominated by Russia, as reflected in Russia's overwhelming budgetary contribution, the predominance of Russian staff, and the very limited non-Russian leadership despite the institute's formally broad membership structure.

# 9. Conclusions

This paper shows how the Joint Institute for Nuclear Research (JINR), despite its formal commitment to peaceful science, is structurally embedded in a local ecosystem in the city of Dubna that includes missile producers, drone manufacturers, other sanctioned MIC enterprises and security services. The analysis demonstrates that the scientific infrastructure of JINR cannot be understood in isolation from the broader technological and industrial environment of the city. The main Russian MIC corporate groups were shown to be represented in Dubna and to collaborate with JINR through a variety of institutional channels. Through long-standing educational programs, knowledge-transfer initiatives, shared infrastructure, and personnel mobility, JINR has contributed to training specialists and developing technologies used in Russia's military campaigns.

The evidence presented throughout this study highlights the importance of the Dubna ecosystem as a highly integrated research-industrial cluster. In this cluster, scientific research, technological development, industrial production, and workforce training operate within a mutually reinforcing institutional framework. Educational programs organized by JINR and partner universities provide a steady flow of highly qualified specialists to enterprises connected with the Russian military-industrial complex. At the same time, knowledge-transfer initiatives and joint technological projects allow innovations generated within fundamental research laboratories to be translated into applied technologies with potential military relevance. These processes illustrate how the boundaries between civilian scientific research and defense-related technological development remain highly permeable within the Russian innovation system.

JINR has also collaborated with impostor and stolen Ukrainian scientific institutions in occupied territories, thereby legitimizing Russia's occupation and exposing international partners to legal and ethical risks. By integrating these institutions into its research networks, the institute contributes to the normalization of Russia's control over Ukrainian territories within international scientific collaboration frameworks. Such activities extend the political implications of JINR's operations beyond the Dubna region and demonstrate how scientific institutions can become instruments of broader geopolitical strategies.

Despite these activities, JINR continues to maintain broad international collaborations, including access to CERN facilities and global computing infrastructure, raising concerns about the indirect support of Russia's military capabilities through international scientific cooperation as shown in Figure 3. After the United States restricted scientific cooperation with Russia in 2014, JINR redirected much of its research activity to European institutions, particularly CERN. Its international standing and limited scrutiny of its ties to Russia's MIC and the FSB allowed it to preserve cooperation with European organizations after 2022, often substituting sanctioned Russian institutions. This enabled JINR scientists to retain access to advanced technologies that could be transferred to military and security actors through Dubna's tightly connected research-industrial ecosystem.

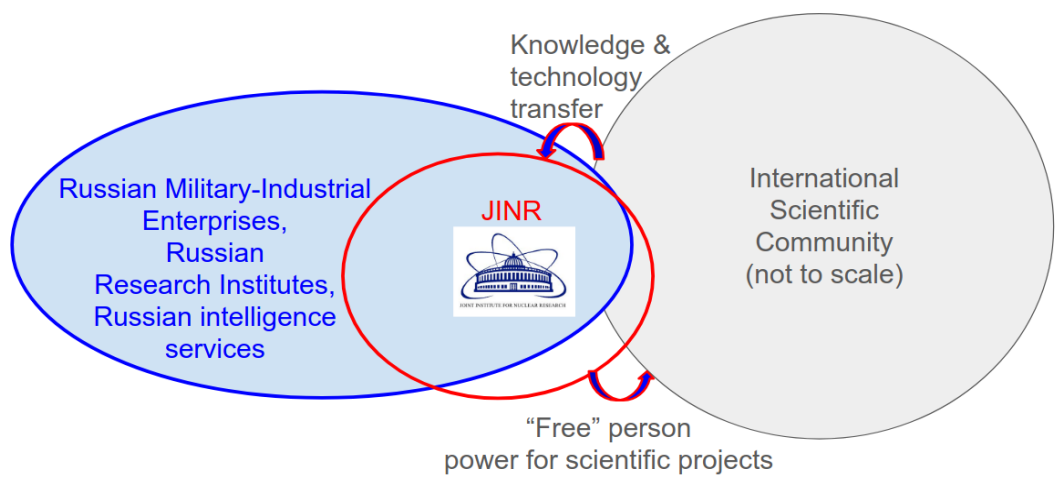

**Fig. 3.** Continued collaboration with JINR enables training and technology transfer, supporting Russian military R&D and production while circumventing existing sanctions.

The persistence of these collaborations illustrates a broader structural challenge in the governance of international scientific cooperation. Large scientific infrastructures are typically designed to promote open collaboration and the global circulation of knowledge. However, in contexts where scientific institutions are embedded in national technological and military systems, such openness may create unintended channels through which advanced technologies and expertise can circulate into defense-related sectors. The case of JINR therefore highlights the dual-use nature of many contemporary scientific infrastructures and the difficulties involved in regulating them through existing sanctions regimes.

The JINR case also demonstrates how international scientific partnerships can indirectly contribute to the resilience of sanctioned technological ecosystems. By maintaining access to global research networks, computational infrastructure, and advanced instrumentation, Russian research institutions can continue to participate in cutting-edge scientific projects while simultaneously sustaining technological capabilities relevant to the military-industrial complex. This dynamic complicates traditional approaches to sanctions policy, which often focus on direct technology transfers while underestimating the role of collaborative scientific networks in maintaining technological capacity.

The JINR case illustrates how Russian scientific research institutions are used to circumvent sanctions, underscoring the need for coordinated enforcement among Ukraine, the EU, and the G7, as well as greater awareness within the international scientific community. More broadly, the JINR case illustrates how international scientific infrastructures may inadvertently contribute to the resilience of sanctioned technological ecosystems. Addressing these challenges will require not only more effective sanctions coordination but also a deeper reflection within the global scientific community on the governance of international research collaborations in contexts of geopolitical conflict.

# List of References